# Software Repositories and Machine Learning Research in Cyber Security

**Mounika Vanamala and Keith Bryant, Alex Caravella**
*Department of Computer Science, University of Wisconsin-Eau Claire, United States*

Corresponding Author:
Mounika Vanamala
Department of Computer Science, University of Wisconsin-Eau Claire, United States
Email: vanamalm@uwec.edu

**Abstract:** In today's rapidly evolving technological landscape and advanced software development, the rise in cyber security attacks has become a pressing concern. The integration of robust cyber security defenses has become essential across all phases of software development. It holds particular significance in identifying critical cyber security vulnerabilities at the initial stages of the software development life cycle, notably during the requirement phase. Through the utilization of cyber security repositories like The Common Attack Pattern Enumeration and Classification (CAPEC) from MITRE and the Common Vulnerabilities and Exposures (CVE) databases, attempts have been made to leverage topic modeling and machine learning for the detection of these early-stage vulnerabilities in the software requirements process. Past research themes have returned successful outcomes in attempting to automate vulnerability identification for software developers, employing a mixture of unsupervised machine learning methodologies such as LDA and topic modeling. Looking ahead, in our pursuit to improve automation and establish connections between software requirements and vulnerabilities, our strategy entails adopting a variety of supervised machine learning techniques. This array encompasses Support Vector Machines (SVM), Naïve Bayes, random forest, neural networking and eventually transitioning into deep learning for our investigation. In the face of the escalating complexity of cyber security, the question of whether machine learning can enhance the identification of vulnerabilities in diverse software development scenarios is a paramount consideration, offering crucial assistance to software developers in developing secure software.

**Keywords:** Machine Learning, Topic Modeling, Cyber Security, CAPEC, MITRE

## Introduction

Cyber Security is a field that implements strategies to protect networks, devices, and data against unauthorized access. It involves maintaining the confidentiality, integrity, and availability of digital information (CISA, 2021). An alarming increase of 600% in cybercrime has been witnessed since January 2020. Small businesses have received a notable percentage of the attacks, with a 400% upswing during this period. Projections indicate that cyber security breaches could result in damages totaling a staggering $10.5 trillion by 2030. Attackers exploit vulnerabilities and errors in software, firmware, and hardware. Software repositories play a vital role in identifying, defining, and cataloging public cybersecurity vulnerabilities. Many resources, including the National Institute of Standards and Technology (NIST) cyber security framework, Common Attack Pattern Enumeration and Classification (CAPEC), Common Vulnerabilities and Exposures (CVE), and Common Weakness Enumeration (CWE) are available. These repositories require continuous updates to aid software against newly identified threats. During the requirements phase of the software development process, the identification of relevant cybersecurity vulnerabilities can significantly reduce time expenditure and improve protection against potential threats.

Incorporating measures during the initial coding stage can minimize the need for extensive recoding to combat vulnerability concerns. With the increasing complexity of cyber security, the role of Machine Learning (ML) in streamlining vulnerability detection during different software development phases becomes increasingly important. This holds particularly true in facilitating software developers and organizations to create robust and secure software right from the requirements phase. The automation of identifying critical vulnerabilities within the software is a primary objective. By automating this process, the potential for human error is reduced, as well as labor intensity. This, in turn, translates to financial savings and the protection of sensitive data.

In this study, researchers explore ML algorithms and their application, primarily potential, in cyber security and the fields related to it.

Software repositories play a vital role by offering insights into vulnerabilities and strategies to enhance cyber defense. These repositories often intersect and store data in varying formats. Numerous organizations, both private and governmental, maintain their own software repositories, often drawing from public ones like CAPEC or The National Institute of Standards and Technology (NIST).

MITRE is a non-profit entity that operates to promote technological and national security progress. As a federally funded organization, MITRE collaborates with various US government organizations. Currently, MITRE oversees more than two hundred distinct projects in addition to managing CAPEC, CVE, CWE MITRE ATT, and CK.

The CAPEC catalog serves as a repository for classifying and arranging cyber security attack patterns that are accessible to the public. These patterns delineate the methods cybercriminals employ to exploit both novel and existing vulnerabilities in software. Professionals such as analysts, developers, testers, and educators employ CAPEC to enhance comprehension of cyber threats and strengthen defense measures (MITRE, 2022). Two large entities utilizing CAPEC as a foundational resource for software security are IBM and Computer Aided Integration of Requirements and Information Security (CIARIS) (MITRE, 2022). Numerous software choices exist for scanning programs to identify vulnerabilities, drawing upon CAPEC or a combination of other repositories.

The Common Vulnerabilities and Exposures (CVE) program is responsible for identifying, defining, and documenting publicly available cyber security vulnerabilities. CVE operates under the sponsorship of the U.S. Department of Homeland Security (DHS) and the Cyber Security and Infrastructure Security Agency (CISA), as outlined on the cve.org website. The repository currently hosts 187,544 accessible CVE records. This repository serves as a compilation of vulnerability incidents (records) that are analyzed and validated by the CVE Numbering Authority (CNA). Each CVE entry encompasses comprehensive details about the attack incidents, with sections including name, status, description, references, phase, votes, and comments. These validated vulnerabilities represent distinct occurrences valuable for research, particularly for the comparison of different threat types. This dataset proves instrumental in deriving statistics, a practice utilized by organizations like MITRE.

The Common Weakness Enumeration (CWE) serves as a repository managed by the community, encompassing a compilation of relevant vulnerability types in both software and hardware (Kanakogi *et al*., 2022). This searchable CWE repository is openly accessible, featuring more than six hundred weaknesses that can be referenced at no cost to the user. Notable entities like Apple, HP, IBM, and Red Hat contribute to the ongoing enhancement of CWE entries. Presently, the CWE list is presented through various perspectives, including a dictionary, classification tree, graph, and slices by topic. Individuals, companies, and even the U.S. National Vulnerability Database (NVD) turn to CWE for referencing or incorporating into their systems. The conjunction of CWE with CAPEC is commonplace to achieve a more comprehensive understanding of software security defense.

MITRE also offers the Adversarial Tactics, Techniques, and Common Knowledge (ATT and CK) repository. ATT and CK function as an attack matrix, encompassing fourteen distinct categories of tactics. These tactic categories feature explanations of attack forms, strategies for mitigation and detection, as well as supplementary sub-techniques that are important to defensive methods. The ATT and CK knowledge repository includes profiles of notable cyber threat groups, a valuable resource in the event of a cyberattack. These attacker profiles are referenced for understanding the patterns of attacks, as well as the tactics employed. ATT and CK have a variety of applications and there are variations in different platform types (e.g., Windows, Linux, cloud, etc.).

NIST is a government agency responsible for devising technology, metrics, and standards to foster innovation and strengthen economic competitiveness among organizations based in the United States (NIST, 2022). NIST's recommended guidelines encompass sets of security controls tailored for information systems, which have government endorsement and encompass optimal security practices used across various industries. NIST extends these guidelines to information systems, which also have government backing and contain security practices spanning diverse sectors. A great example of NIST's standards is the NIST cyber security Framework, designed to enhance risk management for businesses that adopt this model. It is not universally applicable, but this framework has effectively countered and alleviated numerous risks faced by organizations, contributing to its success and adoption.

This study represents an extension of prior research and educational endeavors aimed at creating an automated recommendation system for pinpointing highly correlated CAPEC attack patterns present in a Software Requirements Specifications (SRS) document (Vanamala *et al*., 2020a). The groundwork for this project stems from earlier research, including "topic modeling and classification of CVE database" and "analyzing CVE database using unsupervised topic modeling" (Vanamala *et al*., 2020b).

In a prior study conducted in 2019, research focused on Common Vulnerabilities and Exposures (CVEs) sourced from the national vulnerability database. This investigation employed Topic Modeling techniques (Mounika *et al*., 2019). Specifically, a Gensim Latent Direct Allocation (LDA) was the chosen topic model, while the visualization of these topics was facilitated by pyLDAvis. The research process encompassed multiple steps, including data processing, text cleaning, bigram, and trigram models, topic model construction, visualization of topics and their corresponding keywords, and finally, the association of topics with the Open Web Application Security Project (OWASP) top 10 list.

OWASP serves as a non-profit organization supplying practical insights into application security, catering to the needs of both security professionals and developers. The research utilized a dataset of 121,716 CVE entries, spanning the years from 1999-2019. The topic model was instrumental in identifying trends in 5-year increments. These trends were then evaluated against 5-year periods from the OWASP top 10. Remarkably, the topic model's outcomes exhibited significant similarity with the OWASP top-10 recommendations during the same timeframes. This alignment highlighted the topic model's impressive accuracy compared to the expert analysis conducted by OWASP.

A follow-up study titled "topic modeling and classification of the CVE database" was carried out (Vanamala *et al*., 2020b). Building upon prior research, this study leveraged Latent Direct Allocation (LDA) to identify correlated topics between CVEs and the OWASP top-10. In contrast to previous work, this research introduced a statistical analysis component using standard deviation and coefficient of variance to compare the outcomes of both lists.

Through the application of this statistical analysis, the results of the manual mapping (from the earlier project) and the automated mapping (in this project) exhibited remarkably similar values. These findings indicated that the automated mapping process successfully approximated the manual mapping process employed in the initial project. Notably, the creation of an automated mapping mechanism for comparing two substantial datasets was considered a success. This machine learning model effectively emulated the outcomes of expert analysis, showcasing its capacity to reproduce these outcomes.

Acknowledging the extensive CAPEC database, the study "recommending attack patterns for software requirements document" was conceived (Vanamala *et al*., 2020a). This study transitioned from focusing on CVE data to CAPEC to offer more practical assistance to software developers during the Requirements Engineering (RE) phase. This shift aimed to enhance cyber security support during software development.

The study employed an automated mapping approach to create topic models for the Software Requirements Specification (SRS) document and CAPEC, building on the methodology from the previous project. An SRS document outlines the intended functionalities and outcomes of a proposed software program. Identifying cyber security vulnerabilities from the SRS document allows for the incorporation of cyber security standards into the software development process. However, manually parsing through the extensive list of 555 CAPEC attack patterns to identify the most relevant ones demands considerable time and expertise. The introduction of automation via Latent Direct Allocation (LDA) accelerates this process, minimizing human error and reducing time constraints.

The metric used to gauge the similarity of topic distribution between keywords from the SRSs and those from CAPEC was cosine similarity. This metric's ability to account for document size variation made it a great choice in this context. In summary, the topic model effectively identified relevant CAPEC attack patterns present in the SRS document.

The research also entailed reviewing recent and related literature, focusing on research topics within the realms of cyber security and Machine Learning (ML). This assessment encompassed evaluating the types of ML utilized, the metrics employed, and their relevance to cyber security. The aim was to identify novel processes or insights that could further enhance the ongoing research.

Research was undertaken to establish connections between the Common Vulnerabilities and Exposures (CVE) and Common Attack Pattern Enumeration and Classification (CAPEC) databases (Kanakogi *et al*., 2022). The primary objective was to enhance the comprehensiveness of the cyber security repository by autonomously linking these two databases using Artificial Intelligence (AI). While CVE offers detailed insights into specific attack incidents, CAPEC categorizes attack patterns. While valuable in isolation, a database that interconnects both resources would substantially enhance understanding of attack types by linking attack categories with concrete attack instances.

Researchers embarked on the task of linking CAPEC-IDs to CVE-IDs, employing three algorithms: Sentence BERT (SBERT), Term Frequency-Inverse Document Frequency (TF-IDF), and Universal Sentence Encoder (USE). These algorithms were evaluated under three distinct metrics/experimental patterns: Document, per section, and section average. An exception was noted for SBERT and document due to token limitations. Ultimately, the research determined that TF-IDF was the most suitable algorithm across all testing scenarios.

The research closely aligns with our ongoing efforts, particularly in our shared objective of streamlining the connection between cyber security information and end-users through automated processes. The researchers also suggested a future exploration of ensemble learning, a technique that combines results from diverse tests. They anticipate that incorporating a wider array of professional cybersecurity resources would only make the assistance provided more robust.

Another study was centered on developing a template for assessing the security status of established networks. This initiative was designed to offer a checklist that aids chief security officers in identifying inadequate security measures (Alyami *et al.*, 2021). The checklist encompasses thirty questions spanning various aspects including operating systems, anti-bot software, antivirus software, IPS software, application control software, zero-day protection, and DDoS solutions. Additionally, a strategy table is provided to guide the evaluation of concerns and offer support for risk mitigation if identified. This research topic addresses the specific need to evaluate cyber security threats by assisting in assessing the effectiveness of current security measures. Many projects within the realm of cyber security incorporate the assessment of existing protection status. This approach proves highly valuable for maintaining the ongoing security of established systems and networks.

In contrast, our research is geared toward assisting programmers during the initial phases of software creation to enable secure programming before program implementation. While both research areas are vital for cyber security, they focus on different stages of the cyber security lifecycle.

The following research focused on the development of the Security Assessment Model (SAM), outlined by Prasad *et al*. (2022). SAM stands out as a sophisticated and multifaceted framework crafted to evaluate software for security vulnerabilities. The researchers leveraged established references like CWE, ISO/IEC 25010, SMARTS/SMARTERS, and AHP to support their methodology. To gauge its ability, the model's weighted system was put to the test using an extensive dataset comprising 150 widely used open-source JAVA applications procured from GitHub, alongside 1200 test cases from the OWASP benchmark. This extensive dataset played a pivotal role in validating and assessing SAM's operational capabilities. The research outcomes were deemed successful since SAM not only succeeded in categorizing security levels but also in generating a prioritized weighted list tailor-made for developers to tackle. This weighted list eliminates the need for a triage process, resulting in significant time savings. It's noteworthy that this is the only model, at the time of this research, that has created a working priority list of deficiencies for JAVA programs.

### *Machine Learning*

Machine Learning, a facet of Artificial Intelligence (AI), finds application in handling tasks involving extensive data and automating the recognition of meaningful patterns within data (Shalev-Shwartz and Ben-David, 2014). ML tackles tasks that are either too intricate or notably time-consuming for humans. The domain of ML is broadly divided into two categories: Supervised and unsupervised. The key difference between these ML variants is the utilization of labeled or unlabeled data. In supervised ML, labeled data is employed for both training and testing, whereas unsupervised ML lacks a clear distinction between the data used for training and testing.

### *Unsupervised Machine Learning*

Unsupervised learning leverages the power of ML algorithms to group, correlate, and/or reduce the dimensions of vast amounts of data without labels (IBM, 2019). Clustering involves separating data into clusters, aiding in the identification of similarities and differences within the data. The role of unsupervised algorithms lies in establishing or inferring the relationships among these clusters. Dimensional reduction serves as a mechanism to trim down data inputs while maintaining data accuracy. This facilitates precise and expedited outcomes, particularly when dealing with larger datasets.

### *Topic Modeling*

Topic modeling is a machine learning technique designed to ascertain the correlation between topics. primarily applied in the analysis of textual data, it delves into uncovering connections between topics and assessing their significance. This method operates within the realm of unsupervised machine learning and finds utility in the field of cyber security. The process of topic modeling involves extracting prevalent themes from data and subsequently evaluating their importance.

Various approaches have been taken to implement topic modeling, encompassing purely unsupervised methods, purely supervised methods, as well as hybrid methodologies that combine aspects of both.

Topic modeling can extract meaningful topics from datasets and subsequently gauge the importance of these topics. The technique draws upon diverse algorithms for topic extraction, each distinguished by its metrics for assessing the correlation between topics. The entire process of topic modeling algorithms includes two main phases: Data preparation and topic extraction. Depending on the approach, topic modeling can be done through unsupervised methods, supervised methods, or a blend of the two.

The distinguishing factor between unsupervised and supervised topic modeling lies in the vectorization process. In unsupervised topic modeling, the algorithm identifies and extracts topics from untrained topic categories (Krzeszewska *et al*., 2022). On the other hand, supervised learning involves furnishing the algorithm with guided examples for it to imitate and reproduce (Shalev-Shwartz and Ben-David, 2014). This contrasts with unsupervised machine learning, where the algorithm works without the delineation of training and testing data. Prior to delving into either of these computational approaches, proper text preparation is imperative.

In the realm of cyber security, topic modeling has emerged as an asset, primarily due to its ability to handle vast volumes of data and effectively categorize it into clusters for classification purposes. As this methodology's popularity has risen, it has found practical application in the identification of attack patterns and software vulnerabilities. To pinpoint the intended target of an attack, the process involves gathering data from the ongoing attack and submitting it to the topic modeling algorithm, which then scans for relevant terms and keywords associated with the attack. Once this information is collected, it undergoes a series of sequential phases designed to extract meaningful insights. These phases encompass data processing, text cleaning, constructing the core topic model creating bigram and trigram models, visualizing the extracted topics along with their associated keywords, and ultimately aligning the topics with the OWASP Top 10, a well-known list of the ten most critical web application security risks (Mounika *et al*., 2019). This comprehensive approach ensures a thorough analysis, empowering cyber security experts to quickly identify attack strategies and pinpoint potential software vulnerabilities.

In the exploration of topic modeling, researchers have taken into account the influence of polysemy, a linguistic phenomenon where a single word holds multiple meanings. This consideration is vital since accurately categorizing the intended sense of a word holds substantial sway over the outcomes of topic modeling (Zhu *et al*., 2019). To address this challenge, researchers have introduced the concept of Joint Topic Word-embedding (JWT) processing. This model focuses on analyzing entire sentences to infer a more precise understanding of the words. By doing so, it facilitates the correct placement of words into their respective topic categories.

The findings of this research indicate that the JWT approach yields considerably superior results compared to the direct fine-tuning of pre-trained ELMo or BERT algorithms, both of which are widely used in natural language processing. This advancement not only acknowledges the complexities brought about by polysemy but also demonstrates the effectiveness of the proposed model in enhancing the accuracy of topic Modeling outputs.

### *Latent Dirichlet Allocation*

Latent Dirichlet Allocation (LDA) stands as the predominant algorithm for Topic Modeling. Its application largely being in content analysis, serving as a tool for automating the mapping and organization of vast document archives based on underlying themes. Functioning as a latent variable and mixed membership model, LDA enables documents or data to be associated with multiple latent categories concurrently (Mounika *et al*., 2019). This characteristic avoids oversimplifying document classification and allows for the assignment of multiple labels to the data. LDA also considers topic frequencies and leverages word co-occurrence patterns to deduce themes within the data.

In a study by Sharma *et al*., (2022), they employed LDA to uncover trends in research patterns within blockchain technology. The fundamental idea behind blockchain involves the decentralized recording of transactions in databases across networks. For their investigation, the researchers gathered 993 published papers from digital databases including IEEE, Springer, and ACM. The utilization of LDA enabled the extraction of crucial terms and pivotal documents for each topic, which were then semantically mapped. Dominant research topics, as determined by frequency, included "taxonomy and architecture of blockchain," followed closely by "blockchain implementation and integration over various technologies." To enhance the study's scope, an additional step was introduced to address semantic nuances, thereby achieving a more distinct categorization of the outputs generated by LDA.

To assess its usefulness and suitability, an LDA topic model was executed to identify common software defects in radar systems. Through the incorporation of radar software requirements terminology and the application of supplementary weighting, the bag of words dictionary underwent modification, yielding a distinct lexicon tailored for the LDA algorithm before initiating topic analysis. The outcome was a noteworthy 24% enhancement in accuracy for defect classification. The evaluation of accuracy and recall rates was leveraged to quantify the performance boost in an adapted LSA model when compared to its unaltered counterpart. While the challenge of semantics was resolved by adapting the bag of words approach, the researchers did not explore other potential limitations.

Guo and Li (2021) introduced streamed LDA, an LDA algorithm with a distributed architecture. This modified LDA approach effectively addressed the limitations of traditional LDA, specifically in handling evolving topics, producing accurate inference outcomes, and swiftly accommodating new data instances. Their experimentation encompassed seven datasets, including NeurIPS, all-the-news, New York Times (NYTimes), and Public Medicine (PubMed). By comparing it to the SVB-LDA model, the proposed approach demonstrated enhanced topic consistency through repeated assessments, achieved by integrating the optimal log-likelihood value derived from the SVI-LDA model. The researchers optimized inference quality by fine-tuning local updates and step size, resulting in reduced perplexity metrics indicative of improved word generation coherence compared to SVB-LDA. Additionally, latency comparison, gauged by training stage runtimes, favored the proposed model, showcasing consistently lower execution times than the SVB-LDA.

### *Latent Semantic Analysis*

Latent Semantic Analysis (LSA), an alternate approach within topic modeling, stands as a natural language processing algorithm used to uncover connections between documents and their terms (Prakash *et al*., 2022). LSA is frequently applied to extract and portray the contextual significance of words, as well as to compute similarity among words, sentences, or complete documents. This technique acquires topics through matrix decomposition applied to a document-term matrix using Singular Value Decomposition (SVD), a mathematical technique for matrix analysis (Bellaouar *et al*., 2021).

To enhance document summarization, one study focused on augmenting topic retrieval and its semantic associations for refining search query results by employing LSA (Al-Sabahi *et al*., 2018). The utilized datasets encompassed DUC 2002, DUC 2004, and multilingual 2015 single-document summarization. The integration of word embeddings strengthened the LSA model's input matrix weighting strategies. Outcomes showcased enhanced LSA algorithm performance in document summarization because of the proposed model. Evaluation, performed through metrics within the Rogue automatic evaluation package for summarization, highlighted the superior performance of the suggested model compared to two standard LSA models. However, the researchers acknowledged that the model's effectiveness in generating concise document summaries fell short in certain instances.

Ullah *et al*., (2020) conducted another investigation utilizing LSA to assess the usefulness of a teacher/student internet network. The study employed LSA to analyze the semantic alignment between a teacher's questions and students' answers. By automating LSA, the research facilitated the evaluation of student responses through a comparison against predefined answer keywords. The research drew upon data from the LMS of a virtual university, in Pakistan. Remarkably, the model achieved or surpassed human-level accuracy and notably enhanced processing efficiency. The metric employed for comparing student responses to the answer key was the similarity value. Notably, the study's scope was restricted by the absence of an accuracy evaluation metric; however, the researchers have plans to integrate cosine similarity for this purpose in future endeavors.

In the pursuit of enhancing results concerning semantic space, dimensional reduction, and information retrieval, researchers developed a Heterogeneous model (h-LSA) as an advancement over conventional LSA (León-Paredes *et al*., 2017). This h-LSA model undertook the indexing of five thousand text documents sourced from the PubMed Central (PMC) database. The utilization of multi-CPU and GPU architecture notably led to substantially reduced execution times. The model's accuracy was assessed through cosine similarity. Comparative analysis against the standard LSA model revealed execution speeds ranging from 3.82-8.65 times faster. Across three usage scenarios, the h-LSA model showcased superior accuracy in two instances when compared to the standard LSA.

In the year 2020, a research initiative worked to develop and deploy an intelligent decision support system designed to identify plagiarism within software. (Ullah *et al*., 2021). This research integrated a collection of components, including the Term Document Matrix (TDM), the Singular Value Decomposition (SVD) algorithm, Latent Semantic Analysis (LSA), and the Synthetic Minority Over-Sampling Technique (SMOTE) method.

The research process started with tokenization and the determination of term frequencies, following which the TDM was employed to establish Term Frequency-Inverse Document Frequency (TFIDF) weighting. This process allocated local and global rankings to the identified tokens. Following this, the SVD algorithm was employed to decompose the set of weighted matrices generated in the previous stage. LSA was then applied to extract hidden variables from the SVD outcomes. To address the issue of class imbalance during feature training, the researchers utilized the SMOTE method. This technique helped correct instances of inaccurate categorization, enhancing the overall robustness of the system.

These efforts resulted in an impressive classification accuracy of 92.72%, passing the outcomes of five prior research endeavors conducted between 2011 and 2019. Considering their findings, the authors acknowledged a limitation of their approach and outlined a potential avenue for future exploration. The dependency on the SVD inspired researchers to consider the use of LSA, because of the complexities involved with using SVD.

Sanguri *et al*., (2020) put forward the integration of LSA as an enhancement to a pre-existing approach, Document Co-Citation Analysis (DCA), aimed at refining its outcomes. The introduction of LSA brought a notable advantage by infusing semantic similarity into the equation along with a document similarity metric allocated to each document. The research utilized metadata from the Scopus database, specifically focusing on the tourism supply chain domain. By incorporating LSA, the researchers succeeded in expanding the network and cluster analysis between matrices. Restricted data size and time complexity were roadblocks for the researchers.

### *Supervised Machine Learning*

In contrast to unsupervised machine learning, where algorithms analyze and group unlabeled datasets, supervised machine learning utilizes labeled datasets for training algorithms, enabling them to accurately classify data and make predictions. Supervised machine learning is often categorized into two primary problem types. The first pertains to classification problems, in which support vector machines, decision trees, random forests, etc. group, or categorize data into distinct classes. The second category in supervised machine learning comprises regression problems, where algorithms establish relationships between two variables. These models are predominantly utilized for predicting numerical values based on discrete data points.

### *Support Vector Machine*

Support Vector Machines (SVM), also known as SVM, represent an algorithm aimed at identifying an optimal hyperplane within an N-dimensional space capable of classifying multiple data points. The process involves seeking a maximum margin, which corresponds to the greatest distance between data points from two distinct classes. This established hyperplane allows for increased confidence in categorizing future data points situated beyond this maximum margin. Following the identification of these hyperplanes, they serve as decisive boundaries to facilitate the classification of the active data points. The classification is determined based on the side of the plane on which the points fall. Additionally, support vectors play a role alongside the hyperplane, aiding in its orientation and placement. These support vectors contribute to maximizing the margin at which the points are classified (Asim *et al*., 2021).

The SVM algorithm has been applied to predict the upcoming 10-day statistics related to new infection cases, deaths (including anticipated recoveries), and the number of potentially affected individuals. In assessing death rates, new infection cases, and recovery rates over this span, the SVM algorithm exhibited the poorest performance when compared to alternative methods used for score computation across these categories. The R-squared scores obtained were 0.53 for deaths, 0.59 for new infections, and 0.24 for recovery rates. An observed challenge during this study involved the SVM algorithm displaying notably diminished accuracy when presented with a smaller training dataset and computational workload. This research highlights that while SVM did not emerge as the most proficient algorithm for this specific task, its utility lies in the understanding that its accuracy can be notably improved with larger datasets at its disposal, as indicated by Rustam *et al*. (2020).

Another instance pertains to a study conducted by Uddin *et al*. (2019), where SVM was incorporated to retrieve medical articles. This research employed a variety of machine learning algorithms for health predictions. The calculations entailed a comparative analysis between two databases: Scopus and PubMed. The study's findings indicated that SVM was most often employed in conjunction with multiple algorithm types. Furthermore, SVM ranked as the third most frequently utilized algorithm within articles that exclusively employed a single algorithm. The outcomes of this investigation highlight SVM's prevalence as the most utilized algorithm.

A third area of study conducted by Delli and Chang (2018) centered around the application of the SVM algorithm to assess attributes of printing models for real-time defect monitoring. This research aimed to identify instances of defects arising from filament depletion or structural compromise during the printing process. By leveraging SVM, the researchers established two training model categories: "Good" and "bad,". The system examined RGB values within the print to ascertain the presence of defects and if detected, it would halt the printing procedure. Notably, two limitations emerged from this study. Firstly, the printing process required pausing to capture a clear image of the print. Secondly, defects within the vertical plane of the print were not inspected due to the constraints of a single camera. The insights gained from this research can potentially inform our approach to categorizing training models.

Another study employing the SVM algorithm was presented by McAllister *et al*. (2018), focusing on investigating the usefulness of deep feature extraction in the classification of food image datasets. Utilizing SVM, the researchers successfully categorized images into eleven distinct classes, achieving a good accuracy rate of seventy-eight percent. It's worth noting that alongside this classification approach, they frequently combined SVM with complementary methods to enhance outcome precision for specific datasets. In the broader context of food image recognition, their work proved to be successful, with SVM showcasing a range of accuracies, reaching up to approximately ninety-eight percent in certain cases. This research holds potential value for our own investigations, as it demonstrates a diverse range of classification criteria compared to other algorithms that were utilized.

Finally, another study that utilized SVM was conducted by Schrider and Kern (2018), as they integrated machine learning techniques to manage the rising size of population genomic datasets. Their work aimed to distinguish between selective sweeps and neutrality using SVM. The research allowed them to extract and categorize selective sweeps. This approach enabled them to create genomic predictors through the utilization of training sets that were passed.

### *Naïve Bayes*

Naïve Bayes is a statistical classifier employed to assess probabilities related to its target. This approach treats all data and variables as mutually interconnected, determining the likelihood of association with a specific data segment. The probability theorem denoted to is $P(C|X) =(P(X|C) *P(C))/P(X)$. To breakdown this theorem, $P(C|X)$ represents the posterior probability of the target class, $P(C)$ signifies the prior probability of the class, $P(X|C)$ indicates the likelihood of the class predictor and $P(X)$ is the prior probability of the predictor class.

In one study, the Naïve Bayes algorithm was employed to find medical articles and facilitate health prediction through multiple machine learning algorithms. These predictions were conducted by comparing data from two distinct databases, Scopus and PubMed. The research outcomes indicated that, when used in conjunction with other algorithms, Naïve Bayes ranked as the second most frequently employed algorithm. It also emerged as the second most utilized algorithm within articles only relying on a single algorithm. This investigation highlighted Naïve Bayes' effectiveness with sizable datasets and its usefulness in handling reduced training data. Some limitations that were found with it however were that classes were mutually exclusive and there was a presence of dependency between attributes in the classification (Uddin *et al*., 2019).

Research was also conducted leveraging the Naïve Bayes algorithm to gauge the effectiveness of employing deep feature extraction for categorizing food image datasets (McAllister *et al*., 2018). Through the application of Naïve Bayes, they achieved a food recognition rate of ninety-eight percent within images, which was the lowest score of all the algorithms used. Despite Naïve Bayes yielding the lowest percentage in their study, it successfully identified food content within images. This may be relevant for our own research, as Naïve Bayes allows the utilization of continuous values, so we are able to assume a normal distribution with the data.

Another instance where this approach was utilized involves a liver disease study conducted by Rahman *et al*., (2019). The aim was to predict liver disease outcomes efficiently and accurately, with the intention of reducing diagnostic costs in the medical domain. Within the realm of machine learning algorithms, a dataset encompassing 583 liver patient records was examined. During these tests, Naïve Bayes yielded the poorest results, with an accuracy rate of approximately 53%. Similarly, its precision performance scored at 36% and it also demonstrated the lowest effectiveness in logistic regression calculations, coming in at around 53%. Despite Naïve Bayes showing inferior performance in this study, its standout attribute lay in calculating sensitivity within the dataset, where it outperformed all other machine learning techniques. This insight could hold significance for our research, as it highlights Naïve Bayes' capabilities in deriving valuable information from specific datasets.

Naïve Bayes was also used in research aimed at combating academic plagiarism by developing software to assist instructors in identifying plagiarism in students' code, thereby alleviating the time-consuming manual checking process during grading (Ullah *et al*., 2021). The study employed Naïve Bayes as a tool to assess plagiarism within student work. They used four different training sets which were 50, 60, 70, and 80 when they wanted to test the data. The Naïve Bayes algorithm was harnessed to identify similarities after tokenizing and breaking down the code. Limitations encountered in this research primarily centered around challenges in tokenizing data and preparing code breakdowns for similarity detection. This insight can be important to our research, offering insights into how Naïve Bayes can be utilized for similarity detection.

### *Random Forest*

The Random Forest (RF) machine learning algorithm is structured around multiple individual tree data structures. Each tree in the RF has multiple classifications which when combined give the classification of the given data that was input into the RF. The forest employs a process where it selects the trees containing data or classifications most like each data point and then determines the final set conclusion. Once the chosen dataset is assembled for the tree, the algorithm proceeds to randomize the number of attributes, which correspond to nodes and leaves within a standard tree structure. After this, each tree is expanded to its maximum depth and width without any points being removed (Livingston, 2005).

RF offers key advantages in various applications, including reduced risk of overfitting, flexibility, and ease of determining feature importance (Bellaouar *et al*., 2021). In comparison to the decision trees method, RF distinguishes itself by generating numerous trees, effectively lowering overall variance and prediction errors. This technique's flexibility extends to managing both regression and classification tasks. Moreover, RF provides an enhanced capability to assess the significance of different variables, particularly through metrics like Mean Decrease in Impurity (MDI) and Mean Decrease Accuracy (MCA).

However, RF also presents certain challenges that researchers and practitioners should consider. First, its effectiveness relies on sufficient data, particularly as the number of trees (data structures) increases. This can demand a larger volume of data for reliable results. Additionally, the process of creating multiple trees can consume a significant amount of time, making the method computationally intensive. Lastly, interpreting the outcomes of RF can be complex due to the necessity of aggregating results from multiple structures, adding an extra layer of complexity to the result analysis. Despite these challenges, the benefits of reduced overfitting risk, flexibility, and feature importance evaluation make Random Forest a valuable tool in various analytical contexts.

In another study conducted by McAllister *et al*. (2018), the utilization of random trees was explored for the purpose of food item identification in images. The objective was to assess the usefulness of employing deep feature extraction for classifying food image datasets. By employing Random Forest (RF), they achieved an impressive 99% accuracy rate in detecting food items within images, particularly when utilizing deep feature types. In general, the performance of RF was commendable in distinguishing various types of food. However, there were instances, especially with certain datasets, where it encountered challenges in differentiating specific food types, such as small grains. This insight could potentially be valuable in our research endeavors as it might help to reduce result variance and enhance the overall robustness of calculations.

The research conducted by Schrider and Kern (2018) also explored the application of this method in the context of population genetics research. Their aim was to differentiate between various recombination rate classes and highlight specific motifs or sequences present within the gathered genetic data. While the method they initially employed was effective, the researchers managed to identify an alternative approach that offered enhanced computational efficiency and thus, explored the alternative approach.

This research carries valuable insights as it showcases the utility of Random Forest (RF) techniques in identifying distinctive sequences within extensive datasets requiring classification. The study conducted by Schrider and Kern serves as an illustrative example of how RF can be utilized to discover unique genetic patterns among vast volumes of data.

Another notable application of this method is in the realm of medical research, particularly in the study of liver disease. Rahman *et al*. (2019) delved into predicting liver disease outcomes with enhanced efficiency and accuracy, thereby aiming to decrease diagnostic costs within the medical sector. To accomplish this, they utilized machine learning algorithms, specifically the random forest technique.

The study involved a comprehensive analysis of 583 records of patients with liver conditions, forming the foundation for the datasets used in the research. Within this context, random forest emerged as a powerful tool, yielding notable outcomes. The accuracy rate achieved was approximately 74%, accompanied by a precision of 85% and a sensitivity of 81%. These metrics underlined the usefulness of random forests in generating reliable predictions for liver disease outcomes.

Furthermore, the results highlighted the stability of random forest in accurately calculating crucial data points, including true positives, false positives, false negatives, and false positives. This research contributes to the growing body of evidence showcasing the potential of random forest in the medical domain, particularly in optimizing diagnostic processes and enhancing overall medical decision-making.

More Research using this method (Sweeney *et al*., 2022) involves identifying the classification algorithms, feature extraction functions and interplay between classification algorithms and how feature extraction functions impacted the performance of lesion segmentation methods in MRI testing lesion segmentation is the process of segmenting objects within medical related images. Based on the study, they found that RF performed better than other simpler algorithms when calculating the rates of true positive test rates to false positive test rates. From interpretation of the data, it was determined that the model offers little in terms of intuition about classification in general and rather provides computationally complex rules for making predictions (Sweeney *et al*., 2014).

### *Neural Networks and Deep Learning*

Neural networks are sets of mathematical models that simulate the thinking of a human brain. The network's neurons, like the human brain, fire depending upon the input. The mathematical process of a single perceptron,

"neuron" is as follows, summation of the weighted inputs, addition of the bias, and lastly an activation function. The weighted inputs are said to represent the strength of a synapse (i.e., the connection between the perceptrons). A high value indicates high strength and a low value, low strength. Along with this, weights can also be positive or negative values. A negative weight means that there is inhibited neuron activity while a positive weight strengthens that activity (Mohamed, 2017).

One research topic that used neural networks implemented it to find medical articles that used multiple machine learning algorithms to calculate health predictions. To calculate these predictions, they compared two databases called Scopus and PubMed. They found that neural networks were the third most used algorithm when paired with more than one type of algorithm and that it was the most used algorithm with articles that implemented only one algorithm. Based on the results from this research, neural networks were shown to be highly involved in health research. From this, it has been inferred that neural networks benefited due to their advantage in complex nonlinear relations and in application to classification and regression problems. Although they do have these benefits, the limitations found were the 'black box' nature of the algorithm and the computational expense of training the models (Uddin *et al*., 2019).

A second research investigation employed this technique to manage the increasing size of population genomic datasets (Schrider and Kern, 2018). In this study, neural networks were harnessed to proficiently map summary statistics onto parameters, offering computational savings. This method garnered extensive adoption, particularly for its effectiveness in generating outputs for both categorical and continuous parameters. When compared to our research, it is beneficial to see the wide range of data neural networking will be able to benefit when looking at the wide variety of population genetics.

Another research topic that employed neural networking aimed to gauge the use of deep feature extraction for the classification of food image datasets. Employing an adaptable learning rate and conducting one thousand iterations, they achieved a remarkable 99.4% accuracy in identifying food items. This insight could potentially prove beneficial for our own application, as it demonstrates the capability to efficiently iterate through large datasets (McAllister *et al*., 2018).

Utilizing this approach for investigation, a fourth research avenue was explored. It involved the assessment of classification algorithms, feature extraction functions, the interplay between these algorithms, and the impact of feature extraction functions on the efficacy of lesion segmentation methods in MRI testing. The findings of this study revealed that the superior performance of neural networks compared to alternative algorithms can be attributed to their adeptness in establishing decision boundaries. Notably, when dealing with a modest volume of data, performance remains consistent, approximately at 3000 voxels. However, upon surpassing a threshold of roughly 15000 voxels, instability emerges in the unnormalized graph scaling, leading to erratic shifts in scaling values. One notable constraint of employing this methodology, as revealed by the research, is its performance on imbalanced datasets, showcasing subpar results. Insights derived from this study highlight the significance of addressing the challenges posed by unbalanced datasets and developing strategies to mitigate performance issues when applying the algorithm to novel data (Sweeney *et al*., 2022).

## Discussions

Semantics of words have a crucial role in properly categorizing words through ML. Two different words can be processed into the same word, which potentially provides inaccurate classification. One example is the preprocessing of the words desert and deserted, these words both become desert. The meaning of the word deserted is lost. It would be essential for an ML model to be effective in semantic analysis if it were to make recommendations upon relevant vulnerabilities, utilizing the CAPEC database. The next discussion is the consideration of implementing an unsupervised, supervised, or semi-supervised ML model. The goal of this research would be to compare keywords from an SRS document to the keywords of CAPEC vulnerabilities.

Unsupervised Machine Learning (ML) algorithms find their primary utility in tasks involving the segregation of data into clusters, uncovering underlying data relationships, and reducing dimensionality. For instance, dimensionality reduction becomes a valuable tool when dealing with extensive datasets like the large CAPEC dataset, as it aims to streamline data while preserving its integrity.

In the realm of text analysis, research on Latent Dirichlet Allocation (LDA) uncovered substantial adaptation requirements to achieve satisfactory outcomes, largely due to semantic limitations. On the other hand, the Latent Semantic Analysis (LSA) algorithm is designed to capture semantics and establish connections between vectors that words are segmented into. Over time, LSA has frequently been coupled with techniques such as Singular Value Decomposition (SVD) or other intricate algorithms to enhance its effectiveness. It's important to note that evaluating the usefulness of unsupervised methods, in general, can be challenging due to the absence of well-defined metrics to measure model accuracy. This lack of clear metrics adds complexity to the interpretation of results, making it more intricate to discern the quality of outcomes generated by unsupervised ML approaches.

Supervised ML is a less complex process and requires fewer tools than unsupervised ML (IBM, 2019). Supervised ML uses a training dataset and validation techniques to derive accurate results in a timelier manner, compared to unsupervised ML. Unsupervised ML works by clustering objects into like groups, identified by the algorithm. The largest limitation for supervised ML requires obtaining the training data set to prep the implemented algorithm. Supervised ML also is significantly more proficient at obtaining metrics for the accuracy of results.

## Conclusion

Upon recognizing the significance of cyber security vulnerability controls during the software requirement phase, the CAPEC software vulnerability repository emerged as the most practical repository for this study. The arrangement of attack patterns thus facilitates precise identification and seamless referral back to CAPEC for recommended defense strategies. We define and elaborate on topic modeling, as well as unsupervised and supervised ML methods, showcasing recent research instances and the applicability of these approaches. As our research continues, our efforts will involve the implementation of supervised machine learning. The CAPEC repository provides a prelabeled dataset, a valuable asset for training data set implementation. Supervised ML offers the added benefit of proficiently utilizing metrics to fine-tune the ML process, thus enabling thorough evaluation and process enhancement. A training set for the SRS document must either be crafted or located for supervised ML execution. Given the absence of a comparable research framework employing supervised ML, our future endeavors will assess and compare results stemming from Naïve Bayes and RF ML methodologies. Naïve Bayes showcases statistical prowess across both large and small data sets, making it suitable for the modest data set of SRS documents as well as the larger data set encompassing CAPEC Vulnerabilities. RF's capacity to counteract overfitting aligns well with the intricate data from CAPEC. The algorithm returning the most accurate recommendations for CAPEC attack patterns from an SRS document will be harnessed to deploy an automated tool for result processing and visualization.

## Acknowledgment

## Funding Information

## Author’s Contributions

**Keith Bryant and Alex Caravella:** Acquisition of data and analysis and interpretation of data and content written.

**Keith Bryant, Alex Caravella, and Mounika Vanamala:** Conception and design of the article, intellectual content generation, critically reviewed the article.

**Mounika Vanamala:** Contribution to intellectual content ideation and reviewed the article along with the coordination for publication.

## Ethics

This article is original and contains unpublished material. The corresponding author confirms that all of the other authors have read and approved the manuscript and that no ethical issues are involved.